\documentclass[11pt]{article}
\usepackage{epsfig}
\usepackage{amsmath}
\usepackage{color}

\begin{document}
	
	\begin{center}
		{\Large\bf Non-parametric reconstruction of the cosmological \textit{jerk} parameter}
		\\[10mm]
		Purba Mukherjee\footnote{E-mail: pm14ip011@iiserkol.ac.in},
		Narayan Banerjee\footnote{E-mail: narayan@iiserkol.ac.in}\\[5mm]
		
		{\em $^{1,2}$Department of Physical Sciences,~~\\Indian Institute of Science Education and Research Kolkata,\\ Mohanpur, West Bengal 741246, India.}\\[2mm]
	\end{center}
 
\vskip 1.0cm

\begin{abstract}
The cosmological jerk parameter $j$ is reconstructed in a non-parametric way from observational data independent of a fiducial cosmological model. From this kinematical quantity, the equation of state parameter for composite matter distribution is also found out. The result shows that there is a deviation from the $\Lambda$CDM model close to $z=1.5$, at the $3\sigma$ confidence level. 
\end{abstract}

\vskip 1.0cm


\vskip 0.4mm


\vskip 1.0cm

\section{Introduction}
Even after more than a couple of decades of its discovery\cite{perl, riess}, the accelerated expansion of the universe is yet to be attributed to a well-defined matter sector, called the {\it dark energy}, responsible for the alleged  acceleration. Therefore, the quest for dark energy has been alive along all possible ways. A ``reverse engineering'', where one makes an attempt to find the characteristics of the matter distribution from a given evolution history, is amongst the prominent ways for quite a long time. Normally this ``reconstruction'' is related to figure out a physical characteristic of the matter distribution, such as the equation of state parameter of the dark energy $w_{DE}$, or even the potential $V(\phi)$ if the dark energy is taken as a scalar field. \\

Another direction of reconstruction is through the kinematical parameters, such as the deceleration parameter $q = - \frac{1}{aH^2} \frac{d^2 a}{dt^2}$ where $a$ is the scale factor, and $H=\frac{1}{a}\frac{da}{dt}$, the fractional rate of increase in the linear size of the universe called the Hubble parameter. For a long time, $H$ had been the only cosmological parameter which could be estimated from observational data. As $H$ was found to be evolving, the next higher order derivative of $a$, namely $q$  was the quantity of interest, Now that $q$ can be measured and is found to be evolving, the third order derivative of $a$ finds a natural importance. Expressed in a dimensionless way, this quantity called  the ``jerk'' is defined as 

\begin{equation}
\label{jerkdef}
 j = - \frac{1}{aH^3} \frac{d^3 a}{dt^3}. 
\end{equation} 

There has been some work in the reconstruction of a cosmological model through these kinematical parameters. Reconstruction of the deceleration parameter $q$ can be found in the work of Gong and Wang\cite{gong1, gong2}. Reconstruction through the jerk parameter has been carried out by Luongo \cite{luongo43}, Rapetti {\it et al}\cite{rapetti44}, Zhai {\it et al}\cite{zhai45}, Mukherjee and Banerjee \cite{ankan1, ankan2}. Although the possible importance of the jerk parameter in the game of reconstruction was pointed out long back\cite{varun}, not much work has been done to utilize its full potential. Also, the work already done is an estimation of parameters with a functional form of $j$ being used as an ansatz. This is necessarily restrictive, as the functional form for $j$ is already chosen. \\

A more unbiased way is to attempt a non-parametric reconstruction, where the evolution of the relevant quantity is determined  directly from observational data without any ansatz a priori. Such attempts normally involve the reconstruction of $w_{DE}$\cite{sahl1, sahl2, holsclaw1, holsclaw2, holsclaw3, critt, sanjay, zzhang}. However, there is hardly any attempt to model the dark energy through a reconstruction of the jerk parameter in a non-parametric way. Although there is no convincing reason that a reconstruction of kinematic parameters like $q$ or $j$ is more useful than that of a physical quantity like the dark energy equation of state parameter, this indeed provides an alternative route towards the understanding of dark energy in the absence of a convincing physical theory.\\

In the present work, the jerk parameter $j$ is reconstructed for the first time from the observational data in a non-parametric way. We have utilized various combinations of the Supernova distance modulus data, the Cosmic Chronometer (CC) measurements of the Hubble parameter, the Baryon Acoustic Oscillation (BAO) data and also the Cosmic Microwave Background (CMB) Shift parameter data to examine their effect on the reconstruction. \\

The reconstruction yields the result that for most of the combinations, the $\Lambda$CDM model is well allowed within a $2\sigma$ confidence level. For a few combinations however, the $\Lambda$CDM model is not allowed within this level. \\

Indeed there are apprehensions that the CMB Shift parameter data depends crucially on a fiducial cosmological model\cite{elgaroy} and so does the BAO data\cite{carter}. However, we do not ignore them. Our reconstruction is based on the combinations both including and excluding the CMB Shift and the BAO datasets. The final result, when we extract the physical information, that of $w_{eff}$, looks qualitatively very much similar for various combinations of the datasets. \\ 

In section 2, the methodology is discussed in brief and section 3 contains the actual reconstruction. The last section includes a discussion of the results obtained.

\section{The methodology}

At the outset, we do not assume any fiducial model for the universe except that it is given by a spatially flat, isotropic and homogeneous metric given by 

\begin{equation}\label{metric}
ds^2 = -c^2 dt^2 + a^2(t) \left(dr^2 + r^2 d\theta^2 + r^2 \sin^2\theta d\phi^2 \right).
\end{equation}

We pretend that we do not even know the Einstein equations and pick up only the kinematical quantities. We define the reduced Hubble parameter as, $h(z)=\frac{H(z)}{H_0}$. A subscript $0$ indicates the value of the quantity at the present epoch and $z$ is the redshift given as $1+z = \frac{a_0}{a}$. The luminosity distances of any object (such as a Supernova), can be obtained as
\begin{equation} \label{dL}
d_L(z)= \frac{c(1+z)}{H_0} \int_{0}^{z} \frac{dz'}{h(z')},
\end{equation}
For convenience, we define a dimensionless co-moving luminosity distance,
\begin{equation}\label{D}
D(z) \equiv (1+z)^{-1} \frac{H_0}{c} d_L(z).
\end{equation}

Combining Eq. (\ref{dL}) and (\ref{D}) and taking derivative with respect to $z$, we obtain the relation between Hubble parameter and the co-moving luminosity distance as,
\begin{eqnarray} \label{H_from_D}
H(z) &=& \frac{H_0}{D'},\\
h(z) &=& \frac{1}{D'}.
\end{eqnarray} 
where a prime denotes the derivative with respect to $z$. In terms of the dimensionless quantities $h$, $D$ and their derivatives, the jerk parameter can be written as 

\begin{eqnarray} \label{jerk}
j(z) &=&   -1 + \frac {2(1+z)h h' - (1+z)^2 \left(h'^2 + hh''\right)}{h^2} ,\\
&=& -1 + \frac{ (1+z)^2 \left(D' D''' -3 D''^2 \right)  - 2 (1+z) D' D''}{D'^2}. \nonumber
\end{eqnarray}

The uncertainty in $j(z)$, $\sigma_j$ obtained by error propagating Eq. \eqref{jerk} is given below -
\begin{eqnarray}
	\left(\frac{\sigma_j}{j+1}\right)^2 &=& \left\lbrace \frac{2(1+z)\left[ h'\sigma_{h}+ h \sigma_{h'}\right] -(1+z)^2 \left[ 2 h' \sigma_{h'} + h'' \sigma_{h} + h \sigma_{h''} \right]}{2(1+z)hh' - (1+z)^2 (h'^2 + hh'')}\right\rbrace^2  + \left(\frac{2\sigma_{h}}{h}\right)^2 + \nonumber \\ &-& \frac{4(1+z)\left[ h'\sigma_{h}^2+ h \sigma_{h h'}\right] - 2 (1+z)^2 \left[ 2 h' \sigma_{h h'} + h'' \sigma_{h}^2 + h \sigma_{h''h} \right]}{2(1+z)h^2 h' - (1+z)^2 h (h'^2 + hh'')} , \nonumber \\
	 &=& \left\lbrace \frac{(1+z)^2\left[ \sigma_{D'}D''' + D' \sigma_{D'''} -6 D'' \sigma_{D''}\right] -2 (1+z) \left[ D' \sigma_{D''} + D'' \sigma_{D'} \right]}{ (1+z)^2 \left(D' D''' -3 D''^2 \right)  - 2 (1+z) D' D''}\right\rbrace^2  + \left(\frac{2\sigma_{D'}}{D'}\right)^2 + \nonumber \\ &-&  \frac{2(1+z)^2\left[ D'''\sigma_{D'}^2 + D' \sigma_{D''' D'} - 6 D'' \sigma_{D''D'}\right] - 4 (1+z) \left[ D''\sigma_{D'}^2 + D' \sigma_{D''D'} \right]}{ (1+z)^2 \left(D'^2 D''' -3 D'D''^2 \right)  - 2 (1+z) D'^2 D''}. 	
\end{eqnarray}

In order to implement the reconstruction, the widely used Gaussian processes (GP) \cite{rw, mackay, william, gp}, which is a powerful model-independent technique, is adopted. This is a distribution over functions which generalize the idea of a Gaussian distribution for a finite number of quantities to the continuum. Given a set of data points one can use Gaussian processes to reconstruct the most probable underlying continuous function describing the data, and also obtain the associated confidence levels, without assuming a concrete parametrization of the aforesaid function. It requires only a probability on the target function $f(z)$. \\

In cosmology, GP has attracted a wide application in reconstructing or testing models without an apriori fiducial model \cite{1606.04398[24], 1606.04398[25], 1606.04398[26], 1606.04398[27], 1606.04398[28], 1606.04398[29], 1606.04398[30], wang-meng, wang-meng2, zhou-peng, cai-saridakis}. For a pedagogical introduction to GP, we refer to Seikel {\it et al}\cite{1606.04398[25]}. The code developed is publicly available.\\

Assuming the observational data, such as the distance data $D$, or Hubble data $H$, obeys a Gaussian distribution with a mean and variance, the posterior distribution of reconstructed function $f(z)$ can be expressed as a joint Gaussian distribution of different data-sets involving $D$ or $H$. In this process, the key ingredient is the covariance function $k(z, \tilde{z})$ which correlates the values of different $D(z)$ and $H(z)$ at redshift points $z$ and $\tilde{z}$. The covariance function $k(z, \tilde{z})$ depends on a set of hyperparameters (e.g. the characteristic length scale $l$ and the signal variance $\sigma_f$). This approach also provides a robust way to estimate derivatives of the function in a stable manner. The hyperparameter $l$ corresponds roughly to the distance one needs to move in the input space before the function value changes significantly, while $\sigma_f$ describes typical change in the function value. \\

The choice of covariance function, given in (\ref{cov1}) affects the reconstruction to some extent. Here we have used the Mat\'{e}rn ($\nu = \frac{9}{2}$, $p=4$) covariance \cite{rw} between two redshift points separated by $\vert z-\tilde{z} \vert$ distance units, as in equation (\ref{cov2}). This leads to the most reliable and stable results amongst the other significant choices \cite{seikel2013}.

\begin{eqnarray}
\label{cov1}
 k_{\nu=p+\frac{1}{2}}(z,\tilde{z}) = \sigma_f^2 \exp \left( \frac{-\sqrt{2p+1}}{l} \vert z - \tilde{z} \vert \right) \frac{p!}{(2p)!} \sum_{i=0}^{p} \frac{(p+i)!}{i!(p-i)!} \left( \frac{2\sqrt{2p+1}}{l} \vert z - \tilde{z} \vert \right)^{p-i} ,
\end{eqnarray}

\begin{eqnarray}
\label{cov2}
k_{\frac{9}{2}}(z,\tilde{z}) = \sigma_f^2 \exp \left( \frac{-3 \vert z - \tilde{z} \vert}{l} \right) \left[ 1 + \frac{3 \vert z - \tilde{z} \vert}{l}  \frac{27 ( z - \tilde{z})^2}{7l^2} + \frac{18 \vert z - \tilde{z} \vert ^3}{7l^3} + \frac{27 \left( z - \tilde{z} \right)^4}{35l^4}\right] .
\end{eqnarray} 
\begin{figure}[t!]
	\begin{center}
		\includegraphics[angle=0, width=0.6\textwidth]{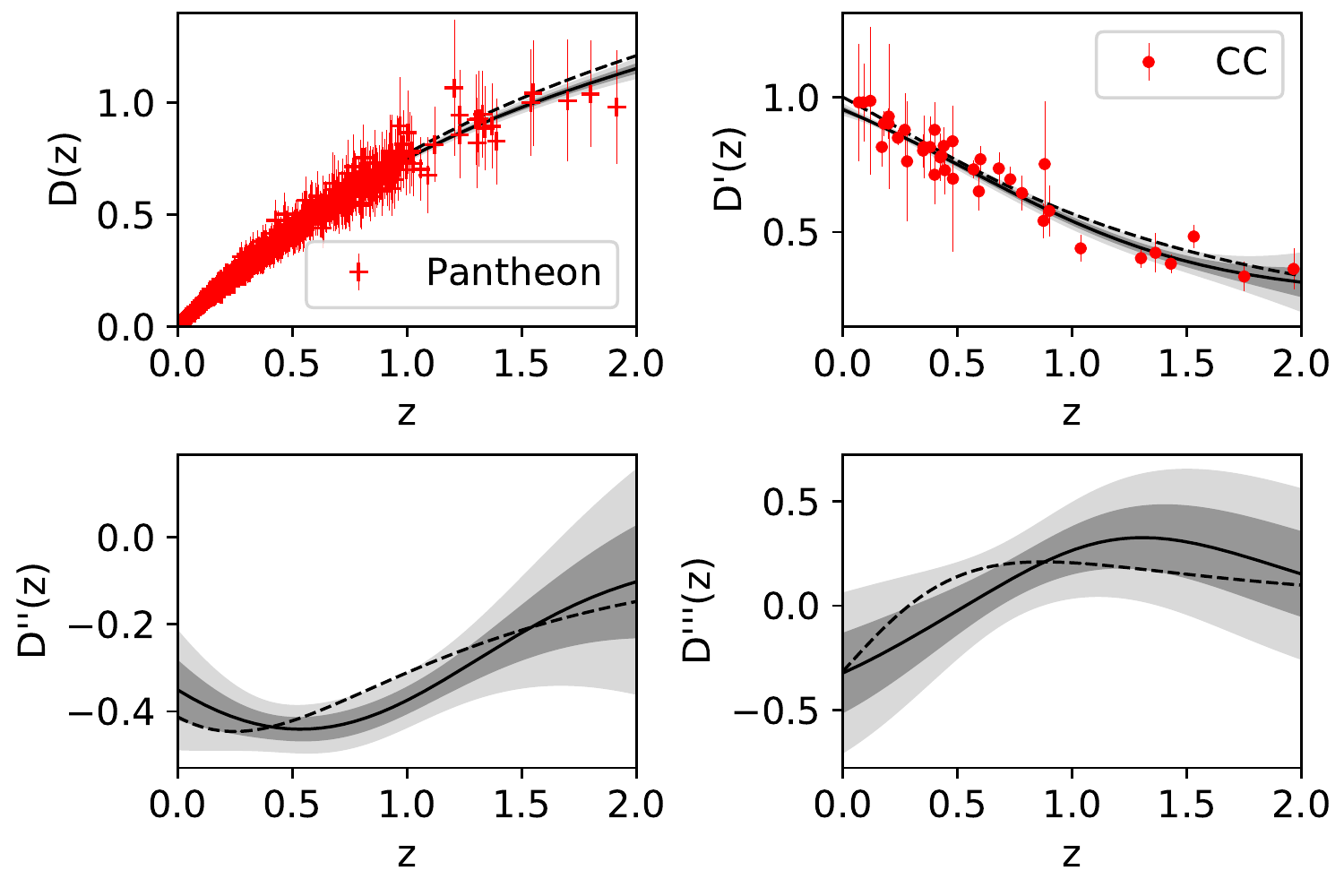}
	\end{center}
	\caption{{\small Plots for the reconstructed dimensionless co-moving luminosity distance $D(z)$, it's derivatives $D'(z)$, $D''(z)$ and $D'''(z)$ using combined Pantheon + CC data  with Planck 2018 best fit prior value $H_0 = 67.27 \pm 0.6 $ km s$^{-1}$ Mpc$^{-1}$ (TT+TE+EE+lowE)\cite{planck}. The black solid line is the best fit curve and the associated 1$\sigma$, 2$\sigma$ confidence regions are shown in grey. The specific points (in the top two figures) with error bars represent the observational data. The black dashed line is for the $\Lambda$CDM model.}}
	\label{all_planck}
\end{figure}

\begin{figure}[t!]
	\begin{center}
		\includegraphics[angle=0, width=0.6\textwidth]{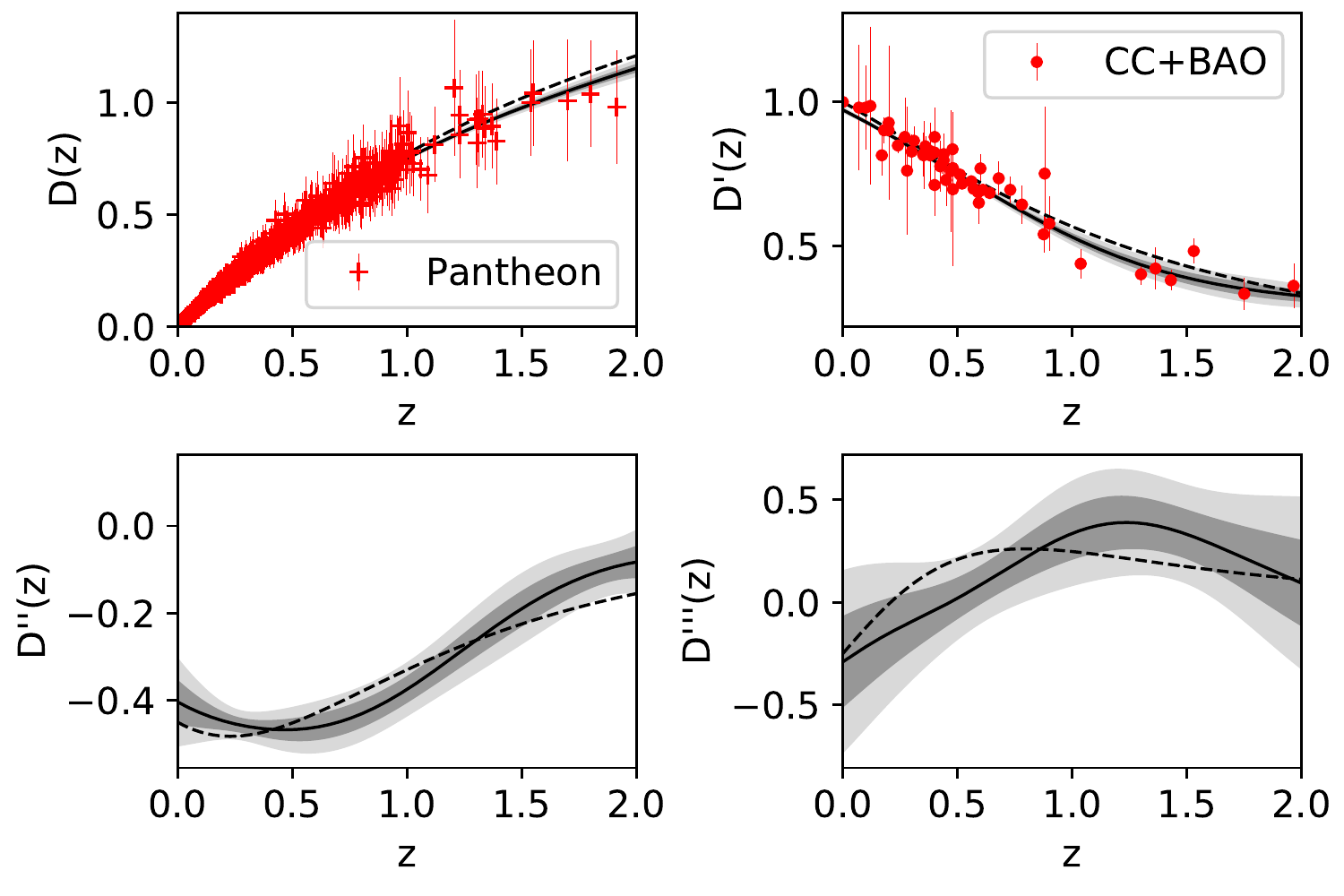}
	\end{center}
	\caption{{\small Plots for the reconstructed dimensionless co-moving luminosity distance $D(z)$, it's derivatives $D'(z)$, $D''(z)$ and $D'''(z)$ using combined Pantheon + CC + BAO + CMB data, with Planck 2018 best fit prior value $H_0 = 67.66 \pm 0.42 $ km s$^{-1}$ Mpc$^{-1}$ (TT+TE+EE+lowE+lensing+BAO) \cite{planck}. The black solid line is the best fit curve and the associated 1$\sigma$, 2$\sigma$ confidence regions are shown in grey. The specific points (in the top two figures) with error bars represent the observational data. The black dashed line is for the $\Lambda$CDM model.}}
	\label{all_planck2}
\end{figure}

\begin{figure}[h!]
	\begin{center}
		\includegraphics[angle=0, width=0.6\textwidth]{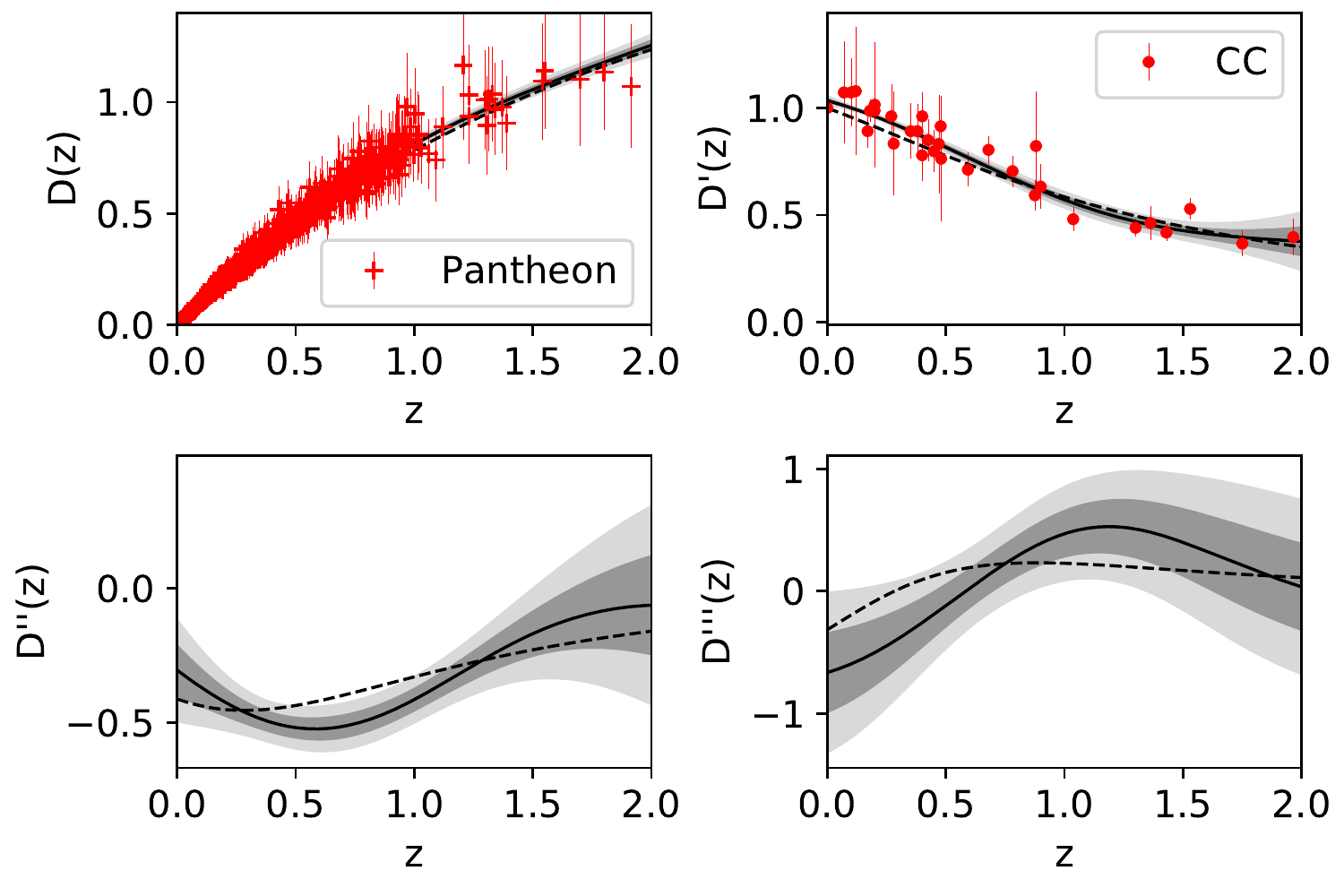}
	\end{center}
	\caption{{\small Plots for the reconstructed dimensionless co-moving luminosity distance $D(z)$, it's derivatives $D'(z)$, $D''(z)$ and $D'''(z)$ using combined Pantheon + CC data, with Riess 2019 best fit prior value $H_0 = 74.03 \pm 1.42 $ km s$^{-1}$ Mpc$^{-1}$ from HST \cite{riess1}. The black solid line is the best fit curve and the associated 1$\sigma$, 2$\sigma$ confidence regions are shown in grey. The specific points (in the top two figures) with error bars represent the observational data. The black dashed line is for the $\Lambda$CDM model.}}
	\label{all_riess}
\end{figure}

\begin{figure}[h!]
	\begin{center}
		\includegraphics[angle=0, width=0.6\textwidth]{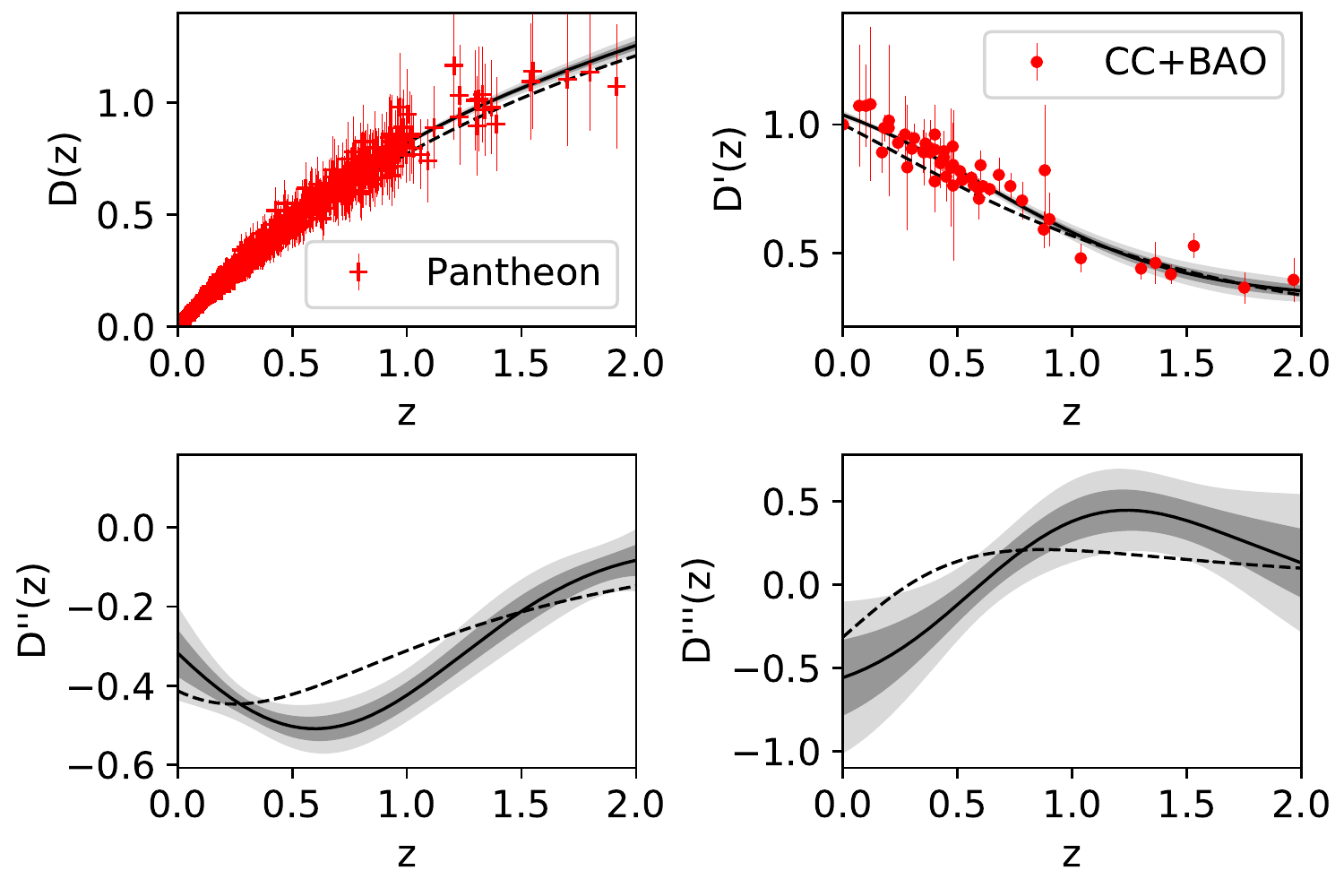}
	\end{center}
	\caption{{\small Plots for the reconstructed dimensionless co-moving luminosity distance $D(z)$, it's derivatives $D'(z)$, $D''(z)$ and $D'''(z)$ using combined Pantheon + CC + BAO + CMB data, with Riess 2019 best fit prior value $H_0 = 74.03 \pm 1.42 $ km s$^{-1}$ Mpc$^{-1}$ from HST \cite{riess1}. The black solid line is the best fit curve and the associated 1$\sigma$, 2$\sigma$ confidence regions are shown in grey. The specific points (in the top two figures) with error bars represent the observational data. The black dashed line is for the $\Lambda$CDM model.}}
	\label{all_riess2}
\end{figure}


The Supernova distance modulus data, Observational measurements of the Hubble parameter, Baryon Acoustic Oscillation data and the CMB Shift Parameter data have been utilized in reconstructing the jerk parameter.\\

We use the $30$ latest $H(z)$ Cosmic Chronometer (CC) data points measured by calculating the differential ages of galaxies \cite{Zhang[61]}, \cite{Zhang[62]}, \cite{Zhang[63]} and the 23 $H(z)$ data points obtained from the radial BAO peaks in the galaxy power spectrum \cite{Zhang[64]}, \cite{Zhang[65]} or the BAO peak using the Ly-$\alpha$ forest of QSOs \cite{Zhang[66]} based on the clustering of galaxies or quasars. One may find that some of the $H(z)$ data points from clustering measurements are correlated since they either belong to the same analysis or there is an overlap between the galaxy samples. Here in this paper, we mainly take the central value and standard deviation of the OHD data into consideration. Therefore, just as in Ref. \cite{geng}, we assume that they are independent measurements. After the preparation of $H (z)$ data, we normalize them to obtain the dimensionless or reduced Hubble parameter $h(z) = H(z)/H_0$. Considering the error of Hubble constant, we calculate the uncertainty in $h(z)$ as,
\begin{equation} \label{sig_h}
{\sigma_{h}}^2 = \frac{{\sigma_H}^2}{ {H_0}^2} + \frac{H^2}{{H_0}^4}{\sigma_{H_0}}^2,
\end{equation} 
where $\sigma_{H_0}$ is the error associated with $H_0$.\\

For the supernova data, we use the Pantheon compilation \cite{pan1}-\cite{pan2} consisting of 1048 SNIa, which is the largest spectroscopically confirmed SNIa sample by now. It consists of different supernovae surveys, including SDSS, SNLS, various low-z samples and some high-z samples from HST. We include the covariance matrix along with systematic errors in our calculation. The distance modulus of each supernova can be estimated as
\begin{equation}
\mu(z) = 5 \log_{10} \frac{d_L(z)}{\mbox{Mpc}} + 25
\end{equation}
where $d_L$ is the luminosity distance in Eq. (\ref{dL}). The distance modulus of SN-Ia can be derived from the observation of light curves through the empirical relation $\mu_{SN} = m^{*}_B + \alpha X_1 - \beta C - M_B$, where $X_1$ and $C$ are the stretch and colour parameters, and $M_B$ is the absolute magnitude. $\alpha$ and $\beta$ are two nuisance parameters. In the Pantheon sample, the corrected apparent magnitude $m_B = m^{*}_B + \alpha X_1 - \beta C$ are reported. Therefore, the colour and stretch corrections are already taken care of in the given dataset. The absolute magnitude of SN-Ia is degenerated with the Hubble parameter, and we fix it to $M_B = -19.35$, the best-fitting value of $\Lambda$CDM. We convert the distance modulus of SN-Ia to the normalized comoving distance through the relation \eqref{D}
\begin{equation} \label{sne_D}
	D(z) \equiv \frac{1}{1+z} \frac{H_0}{c} 10^{\frac{\mu - 25}{5}}.
\end{equation} where $\mu$ is given by the difference between the corrected apparent magnitude $m_B$ and the absolute magnitude $M_B$ in the B-band for SN-Ia.

The total uncertainty or error propagation $\mathbf{\Sigma}_\mu$ and   $\mathbf{\Sigma}_D$ in $\mu$ and $D$ respectively are estimated following the standard practice. The total uncertainty matrix of distance modulus is given by, 
\begin{equation}
	\mathbf{\Sigma}_\mu = \mathbf{C}_{stat} + \mathbf{C}_{sys} 
\end{equation}
where $\mathbf{C}_{stat}$ and $\mathbf{C}_{sys}$ are the statistical  and systematic uncertainties respectively.\\

The uncertainty of $D(z)$ is propagated from that of $\mu$ and $H_0$ using the standard error propagation formula,
\begin{equation}  \label{sne_sigD}
	\mathbf{\Sigma}_D = \mathbf{D}_1 \mathbf{\Sigma}_\mu {\mathbf{D}_1}^T + \sigma_{H_{0}}^2 \mathbf{D}_2 \mathbf{D}_2^{T}
\end{equation}
where $\sigma_{H_0}$ is the uncertainty of Hubble constant, the superscript `$T$' denotes the transpose
of a matrix, $\mathbf{D}_1$ and $\mathbf{D}_2$ are the Jacobian matrices,

\begin{eqnarray}
	\mathbf{D}_1 &=& \mbox{diag}\left(\frac{\ln 10}{5} \mathbf{D}\right) \\
	\mathbf{D}_2 &=& \mbox{diag}\left(\frac{1}{H_0} \mathbf{D}\right)
\end{eqnarray}
where $\mathbf{D}$ is a vector whose components are the normalized comoving distances of all the SN-Ia. \\

The so-called shift parameter is related to the position of the first acoustic peak in the power spectrum anisotropies of the cosmic microwave background (CMB). However the shift parameter $R$ is not directly measurable from the cosmic microwave background, and its value is usually derived from data assuming a spatially flat cosmology with dark matter and cosmological constant.
\begin{equation}
R = \sqrt{\Omega_{m0}}\int_{0}^{z_c}\frac{dz'}{h(z')}
\end{equation} 
where $z_c$ = $1089$ is the redshift of recombination. We use the CMB shift parameter $R = 1.7488 \pm 0.0074$ and matter density parameter $\Omega_{m0} = 0.308 \pm 0.012$ from the Planck's release \cite{planck_cmb} as important supplements of SN-Ia data.\\

In view of the known tussle between the value of $H_0$ as given by the Planck data\cite{planck} and that used prior to the advent of Planck mission, we reconstruct $j$ twice, using both of them separately. The recent global and local measurements of $H_0 = 67.27 \pm 0.60 $ km s$^{-1}$ Mpc$^{-1}$ (TT+TE+EE+lowE), $67.66 \pm 0.42 $ km s$^{-1}$ Mpc$^{-1}$ (TT+TE+EE+lowE+lensing+BAO) with $1\%$ uncertainty (P18)\cite{planck} and $H_0 = 74.03 \pm 1.42 $ km s$^{-1}$ Mpc$^{-1}$ with $2.4\%$ uncertainty (R19)\cite{riess1}, are considered for the purpose. The reconstructed functions $D(z)$, $D'(z)$, $D''(z)$ and $D'''(z)$ are plotted against $z$ for all four sets of the combined datasets, and shown in Fig. \ref{all_planck}, \ref{all_planck2}, \ref{all_riess} and \ref{all_riess2} for the two choices of the prior value of $H_0$. The black solid line is the best fit curve. The shaded regions correspond to the $68\%$ and $95\%$ confidence levels (CL). The true model is expected to lie within the $68\%$ CL. The specific points (in the top two figures in all the four sets) with error bars represent the observational data used in reconstruction. For the Pantheon data, eq. \eqref{sne_D} and \eqref{sne_sigD} are used to estimate the $D$ data points and the uncertainty $\mathbf{\Sigma}_D$ from the observed $\mu$ and $\mathbf{\Sigma}_{\mu}$ respectively. For the CC and BAO data, we consider eq. \eqref{sig_h} and convert the $H$-$\sigma_H$ data to $h$-$\sigma_{h}$ data set. From \eqref{H_from_D} we can clearly see $D'(z)$ is related to $h(z)$. So, we can take into account the $h$ data points, the uncertainty $\sigma_{h}$ associated, and represent is graphically as
\begin{eqnarray}
	D' &=& \frac{1}{h}, \nonumber \\ \vert \sigma_{D'} \vert &=& \frac{1}{h^2} \vert \sigma_{h} \vert . 
\end{eqnarray}
The black dashed line is for the $\Lambda$CDM model. Thus, given a set of observational data points we have used the Gaussian processes to construct the most probable underlying continuous function $D(z)$ describing the data, along with its derivatives $D'(z)$, $D''(z)$ and $D'''(z)$, and have also obtained the associated confidence levels.

\begin{figure*}[t!] 
	\begin{center}
		\includegraphics[angle=0, width=\textwidth]{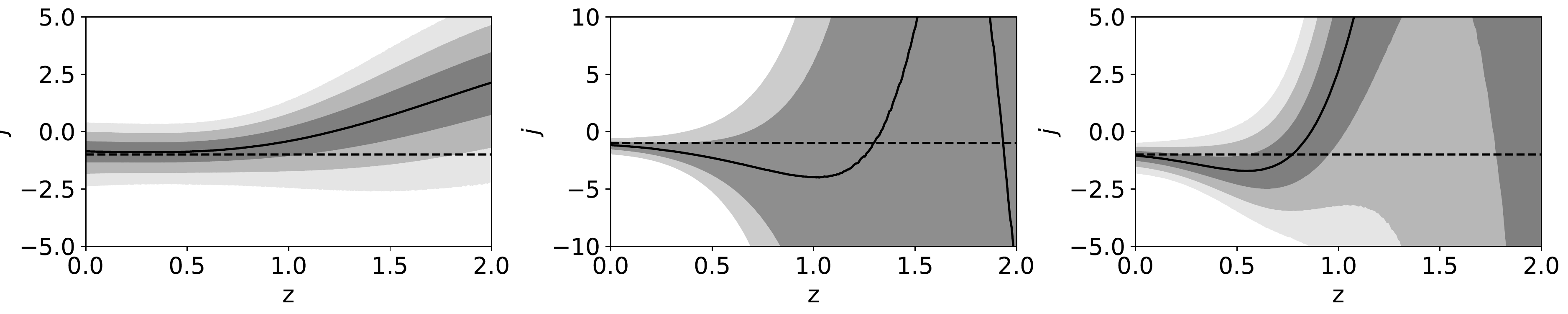}\\
		\includegraphics[angle=0, width=\textwidth]{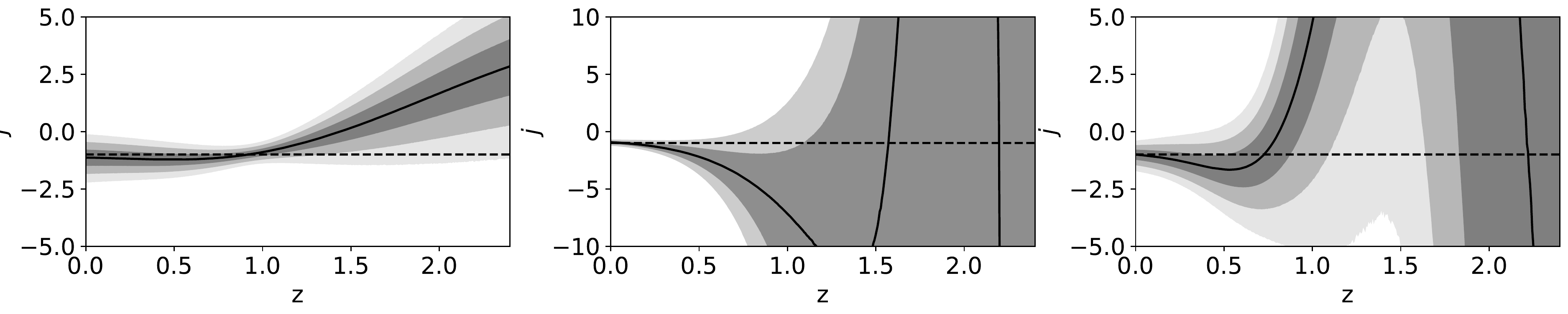}\\
	\end{center}
	\caption{{\small Plots for $j(z)$ reconstructed from different values of the Planck 2018 prior Hubble parameter at present epoch ($H_0 = 67.27 \pm 0.6 $ km s$^{-1}$ Mpc$^{-1}$ \cite{planck} for CC data (left),  Pantheon data (middle), and combined CC+Pantheon data (right) in first row) and, ($H_0 = 67.66 \pm 0.42 $ km s$^{-1}$ Mpc$^{-1}$ \cite{planck} for CC+BAO data (left),  Pantheon+CMB data (middle), and combined CC+BAO+CMB+Pantheon data (right) in second row. The solid black line is the ``best fit'' and the black dashed line represents the $\Lambda$CDM model.}}
\label{jerkplot_p}
\end{figure*}
 
\begin{figure*}[t!] 
	\begin{center}
		\includegraphics[angle=0, width=\textwidth]{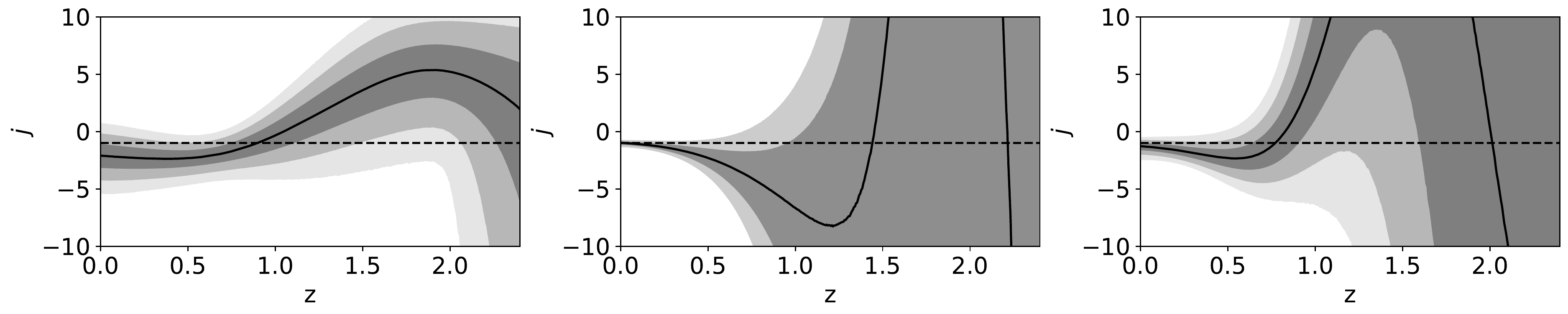}\\
		\includegraphics[angle=0, width=\textwidth]{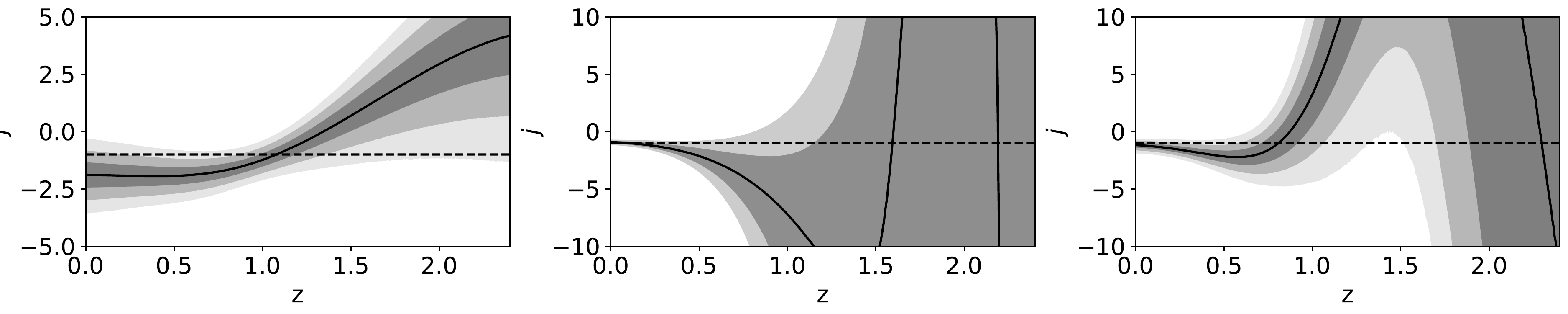}\\
	\end{center}
	\caption{{\small Plots for $j(z)$ reconstructed from different values of the Riess 2019 prior Hubble parameter at present epoch $H_0 = 74.03 \pm 1.42 $ km s$^{-1}$ Mpc$^{-1}$ \cite{riess1} for CC data (left),  Pantheon data (middle), and combined CC+Pantheon data (right) in first row and, CC+BAO data (left),  Pantheon+CMB data (middle), and combined CC+BAO+CMB+Pantheon data (right)) in second row. The solid black line is the ``best fit'' and the black dashed line represents the $\Lambda$CDM model.}}
	\label{jerkplot_r}
\end{figure*}

\section{The reconstruction}

We now reconstruct the cosmological jerk parameter $j(z)$ using the Gaussian Process from the reconstructed function $D(z)$ and its higher order derivatives ($D'(z)$, $D''(z)$ and $D'''(z)$) using eq. \ref{jerk}. Results for the reconstructed jerk is given in Fig. \ref{jerkplot_p} and \ref{jerkplot_r} respectively. The shaded regions correspond to the $68\%$, $95\%$ and $99.7\%$ confidence levels (CL). The black solid line shows the ``best fit'' values of the reconstructed function. Plot shows that the $\Lambda$CDM model, in most of the combinations, is allowed within a $2\sigma$ error bar. \\

However, for the Planck 2018 $H_0$ prior (Fig\ref{jerkplot_p}), the CC + BAO combination (bottom left) and the CC + BAO + CMB + Pantheon combination (bottom right), the $\Lambda$CDM is allowed only in $3\sigma$ and not $2\sigma$ for a brief period. \\

For the Riess 2019 prior, the CC + BAO combination (bottom left in Fig\ref{jerkplot_r}),  the $\Lambda$CDM model is included only in $3\sigma$ and not in $2\sigma$ for most of the evolution between $z=0$ and $z=2.5$. The bottom right plot of this figure shows that for  the CC + BAO + CMB + Pantheon combination, $\Lambda$CDM is not included even in $3\sigma$ close to $z=1.5$. \\

The plots for the ``best fit value'' (black solid lines) of the jerk parameter indicate that $j$ has an evolution, and also, this evolution may well be non-monotonic. \\

The approximate fitting functions for the reconstructed jerk parameter are given in Eq. \eqref{jerk_result}-\eqref{jerk_result2} and \eqref{jerk_result3}-\eqref{jerk_result4} for two sets of combinations, namely CC + Pantheon and the combination of all the data sets. \\ 

For CC + Pantheon dataset combination:

 \begin{align}	\label{jerk_result}
&j_{\mbox{\tiny P18}}(z) = -0.99995 - 1.61516 ~z + 10.0773 ~z^2 - 86.4326 ~z^3 + 310.932 ~z^4 - 601.198 ~z^5 + \nonumber \\ &\hspace{6.5cm} + 680.92 ~z^6 + - 420.081 ~z^7 + 129.508 ~z^8 - 15.6674 ~z^9,\\ \nonumber \\
&j_{\mbox{\tiny R19}}(z) =  -1.08262 - 0.0678615 ~z - 14.3527 ~z^2 + 66.0917 ~z^3 - 154.661 ~z^4 + 183.143 ~z^5 + \nonumber \\ &\hspace{11cm}- 91.3607 ~z^6 + 15.7326 ~z^7, \label{jerk_result2}
\end{align}

and for CC + Pantheon + BAO + CMB combination:
 \begin{align}	\label{jerk_result3}
&j_{\mbox{\tiny P18}}(z) = -0.99996 - 1.60148 ~z + 20.5976 ~z^2 - 179.3920~ z^3 + 683.416 ~z^4 - 1434.995~ z^5 + \nonumber \\ &\hspace{5cm} + 1769.65 ~z^6 - 1264.88 ~z^7 + 513.19 ~z^8 - 109.89 ~z^9 + 9.666 z^{10},\\ \nonumber \\
&j_{\mbox{\tiny R19}}(z) =  -1.04967 - 2.68 ~z + 28.3061 ~ z^2 - 216.726 ~z^3 + 754.522 ~z^4 - 1479.94 ~z^5 + \nonumber \\ &\hspace{4cm} + 1736.11 ~z^6 - 1197.97 ~z^7 + 473.438 ~z^8 - 99.2284 ~z^9 + 8.56616 ~z^{10}. \label{jerk_result4}
\end{align}

\begin{figure*}[t!]
	\begin{center}
		\includegraphics[angle=0, width=0.4\textwidth]{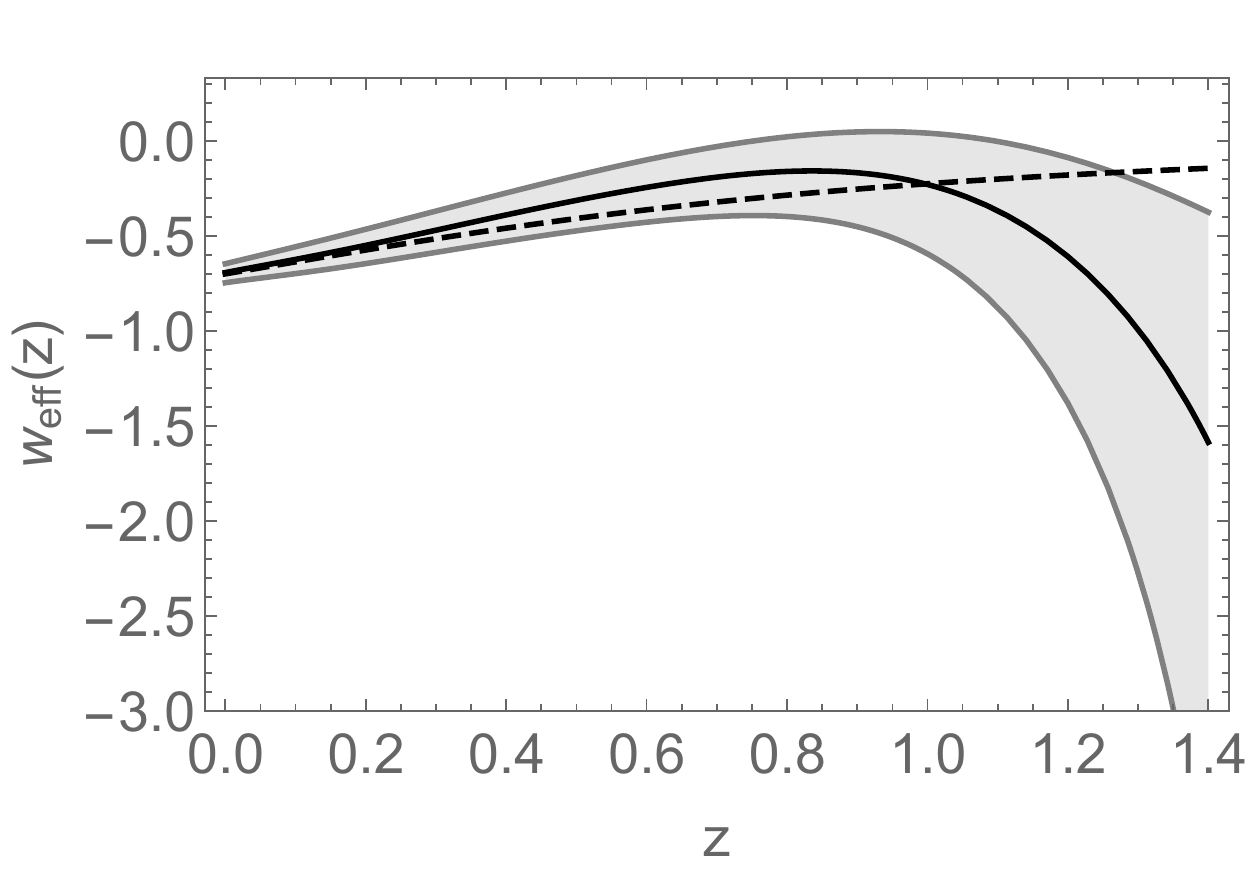} \hspace{1cm}
		\includegraphics[angle=0, width=0.4\textwidth]{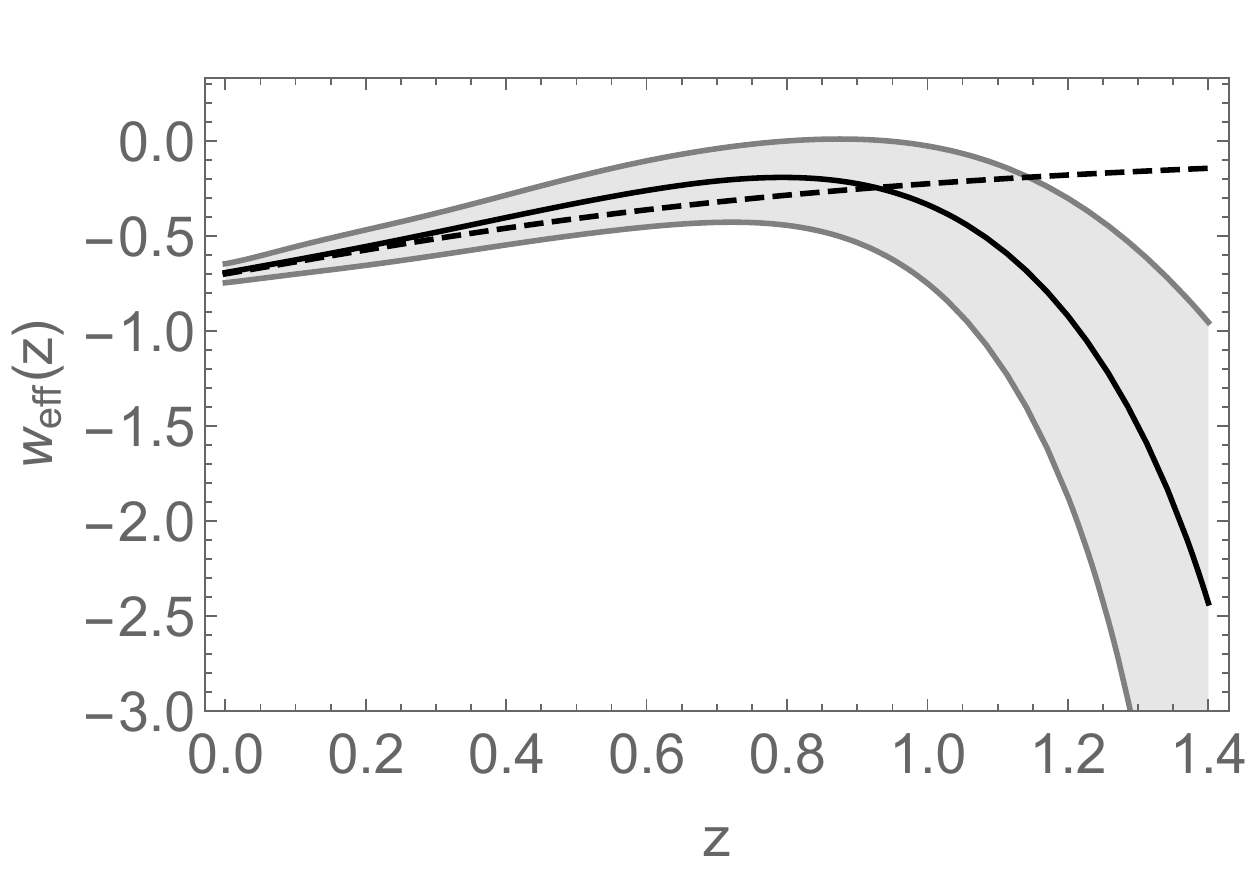}\\
		\includegraphics[angle=0, width=0.4\textwidth]{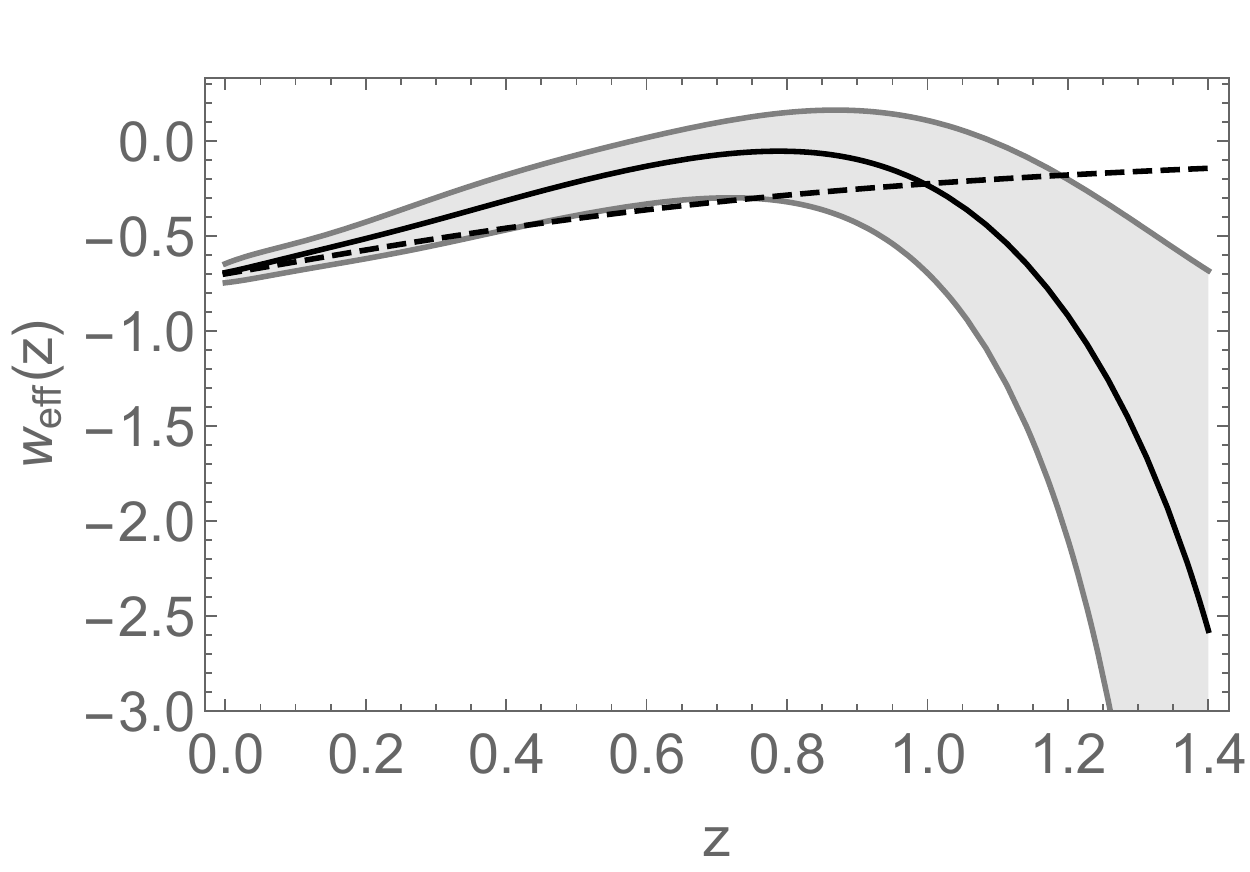} \hspace{1cm}
		\includegraphics[angle=0, width=0.4\textwidth]{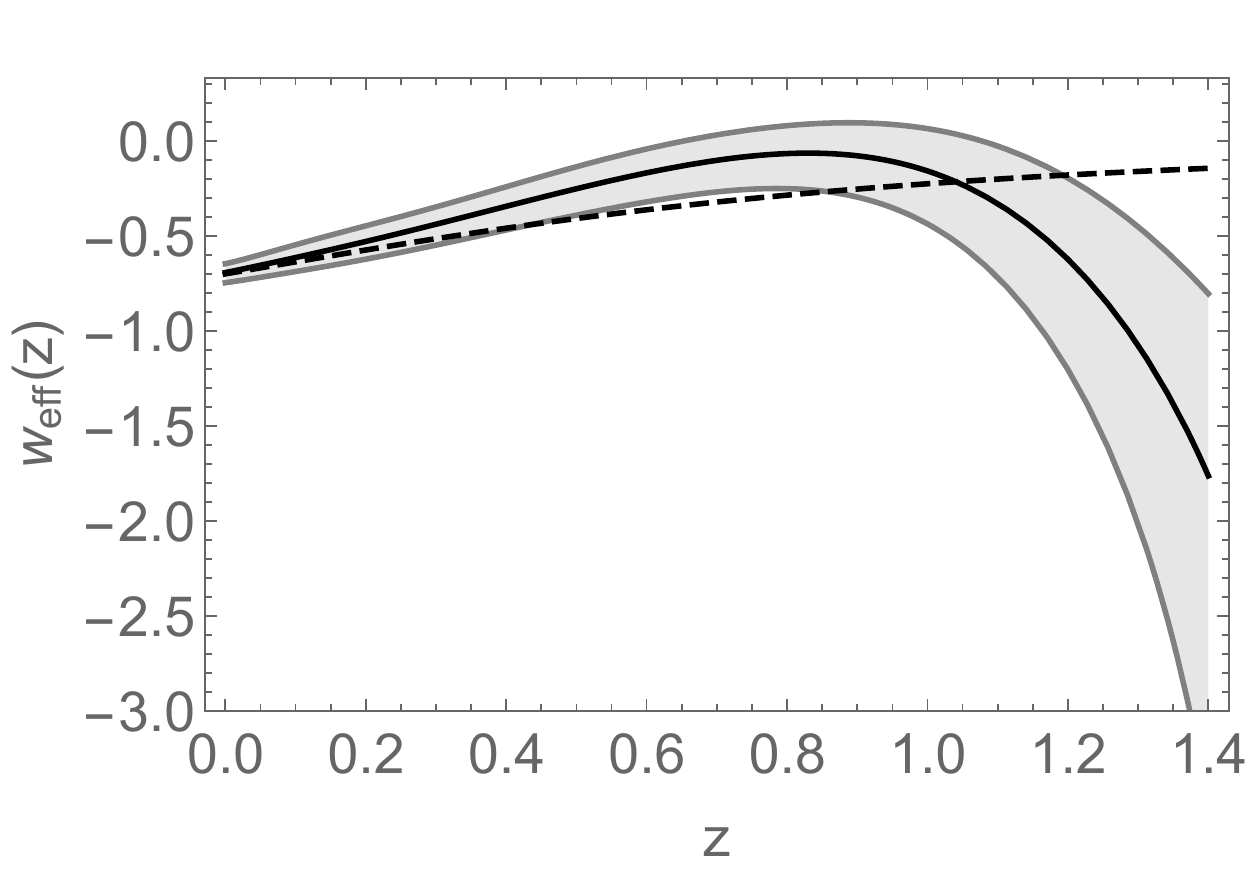}
	\end{center}
	\caption{{\small Plots for the effective equation of state parameter, from the reconstructed jerk $j$, using different prior values of the Hubble parameter at present epoch ($H_0$) for combined CC+Pantheon data (left) and CC+BAO+CMB+Pantheon data (right) using Planck 2018 (top) and Riess 2019 (bottom) data. The black dashed line represents the effective EoS for $\Lambda$CDM model, considering $\Omega_{m0} = 0.308$ \cite{planck_cmb}.}}
	\label{weffplot}
\end{figure*}

We now relax our pretension of not knowing Einstein equations. We use the definition of deceleration parameter
\begin{equation} \label{q}
\frac{\dot{H}}{H^2} = -(1+q),
\end{equation}
in Einstein equations, 
\begin{eqnarray} \label{friedmann}
3H^2 &=& 8\pi G \rho ,\\
2\dot{H} + 3H^2 &=& -8\pi G p ,
\end{eqnarray} where $\rho$ and $p$ are the total energy density and pressure contribution from all components constituting the Universe. Therefore, the effective equation of state parameter is
\begin{equation}  \label{w_eff}
w_{eff} = \frac{p}{\rho} = -\frac{2\dot{H} + 3H^2 }{3H^2 }  = \frac{-1 + 2q}{3}.
\end{equation}

One can write $j$ in terms of $q$ as
\begin{equation} \label{q_from_j}
j(z)= -\left[q(2q+1)+(1+z)\frac{dq}{dz}\right].
\end{equation}
Using equations \eqref{jerk_result}-\eqref{jerk_result4}, for the two datasets, equation \eqref{q_from_j} can be numerically integrated for $q(z)$. For this one has to assume the initial value of the deceleration parameter at the present epoch ($z=0$) i.e., $q_0$. We have chosen $q_0 \simeq -0.54^{+0.07}_{-0.08}$ from reference \cite{q0choice} at $z=0$, and using the solutions for $q=q(z)$ in \eqref{w_eff} we arrive at the effective EoS parameter from reconstructed jerk. We plot the evolution for the effective equation of state parameter in Fig. \ref{weffplot}. The black solid line represent the effective EoS obtained from the reconstructed jerk. The shaded regions show the uncertainty associated with $w_{eff}$ corresponding to the 1$\sigma$ confidence level for the reconstructed jerk parameter (say $j \pm \sigma_j$). The uncertainty in $w_{eff}$ is ascertained by numerically integrating both $j \pm \sigma_j(z)$ alongside $j(z)$ in eq. \eqref{q_from_j} starting from the initial value $q_0$.\\

The approximate functional forms obtained for the effective equation of state parameter are given in Eq. \eqref{weff_result}-\eqref{weff_result2} and \eqref{weff_result3}-\eqref{weff_result4} for two sets of combinations, namely CC + Pantheon and the combination of all the data sets.\\

For CC + Pantheon dataset combination:

\begin{align}	\label{weff_result}
w_{\mbox{\tiny P18}}(z)&= -0.693358 + 0.718592~ z - 15.8327 ~z^{2.67414} + 38.0581 ~z^3 - 86.5953 ~z^4 + 208.338 ~z^5 + \nonumber \\
&+ 518.941 ~z^{7.00109}- 513.523 ~z^8 + 354.204 ~z^9 - 160.4 ~z^{10} + 42.6695 ~z^{11} - 5.05034 ~z^{12} + \nonumber \\ &- 381.065 ~\left| z \right| ^6,\\ \nonumber \\
w_{\mbox{\tiny R18}}(z)&=  -0.692601 + 0.625951 ~z + 3.33012 ~z^{1.81994} - 44.8849 ~z^{3} + 262.946 ~z^{4} - 844.011 ~z^{5} + \nonumber \\
& + 1661.12 ~z^{6} - 2098.3 ~z^7 - 872.5 ~z^9 + 254.023 ~z^{10} - 32.3061 ~z^{11} + 1710.42 ~\left| z \right|^8 , \label{weff_result2}
\end{align}

and, for CC + Pantheon + BAO + CMB combination:
\begin{align}	\label{weff_result3}
w_{\mbox{\tiny P18}}(z)&= -0.69368 + 0.669596 ~z - 0.119192 ~z^2 + 1.4935 ~z^3 - 2.32544 ~z^4 - 0.119214 ~z^5 + \nonumber \\ &+1.57755 ~z^6 + 0.581673 ~z^7 - 1.14239 ~z^8 - 
1.44675 ~z^9 - 0.07217 ~z^{10} + 1.42208 ~z^{11} + \nonumber \\ &+ 1.17954 ~z^{12} - 0.855902 ~z^{13} - 1.74517 ~z^{14} + 1.6586 ~z^{15} - 0.398383 ~z^{16} ,\\ \nonumber \\
w_{\mbox{\tiny R19}}(z)&=  -0.693191 + 0.72215 ~z + 1.38516 ~z^{2.00893} - 11.4426 ~z^3 + 62.3137 ~z^4 - 195.948 ~z^5 + \nonumber \\ &+ 379.509 ~z^6 - 473.887 ~z^7 + 
382.535 ~z^8 - 193.629 ~z^9 + 56.1131 ~z^{10} - 7.13564 ~z^{11} \label{weff_result4}
\end{align}

The value of $w_{eff}$ at $z=0$ is $-0.693^{+0.07}_{-0.08}$ (this depends on the chosen value of $q_0$).  Considering the value of $\Omega_{m0} = 0.308 \pm 0.012$ from Planck data release \cite{planck_cmb}, we can calculate the value of $w_{eff,\Lambda0}$ to be $-0.692$ with $\pm 0.027$ uncertainty at $z=0$ using the standard error propagation method. For higher redshift $z>1.2$, the reconstructed $w_{eff}$ in the present work clearly shows a sizeable departure from the corresponding $w_{eff,\Lambda}$ values of the $\Lambda$CDM model, which can be obtained from \eqref{w_eff}) as,
\begin{equation}
	w_{eff,\Lambda} = -\frac{1}{1 + \frac{\Omega_{m0}}{1-\Omega_{m0}}(1+z)^3}.
\end{equation} 

It should also be mentioned that the nature of $w_{eff}$ as shown in Fig. \ref{weffplot} does not depend critically on small changes in the chosen value of $w_{eff}$ at $z=0$. So we did not include other choices in the figure. 

\section{Discussion}

A reconstruction of the cosmological jerk parameter $j$ is attempted in this work. The reconstruction is non-parametric, so $j$ is unbiased to assume any particular functional form to start with. Also, it does not depend on the theory of gravity, only except the assumption that the universe is described by a 4 dimensional spacetime geometry and it is spatially flat, homogeneous and isotropic. It deserves mention that although a non-parametric reconstruction is there in the literature for quite some time now for reconstructing physical quantities like the equation of state parameter or the quintessence potential, it has hardly been used to reconstruct the jerk parameter. \\

Kinematical quantities, that can be defined with the metric alone (namely the scale factor $a$), form the starting quantities of interest in the present case. As the deceleration parameter $q$ is now an observed quantity and is found to evolve, the next higher order derivative, the jerk parameter is the focus of attention. Surely the parameters made out of even higher derivatives like snap (4th order derivative of $a$), crack (5th order derivative) etc. could well be evolving\cite{capoz}. But we focus on $j$ which is the evolution of $q$, the highest order derivative that is an observationally measured quantity. For a parametric reconstruction of $j$, one can still start from the higher order derivatives\cite{cald, stacho} and integrate back to $j$, and estimate the parameters, coming in as constants of integration, with the help of data. But this does not form the content of the present work as already mentioned. \\

It is found that for various combinations of datasets, the $\Lambda$CDM model is normally included in the $2\sigma$ confidence level. For some combinations, this is included in the $3\sigma$ confidence level but not in $2\sigma$. The most significant departure is for the CC + BAO + CMB + Pantheon combination with the Riess 2019 prior for $H_0$, where the $\Lambda$CDM is not even included in $3\sigma$ for a brief period close to $z=1.5$.  The plots also show that the nature of $j$ does not substantially change for the $H_0$ prior chosen. \\

The polynomials for the best fit curve for $j$ have been worked out. This is done for two combinations, namely CC + Pantheon, where BAO and CMB Shift data are avoided for reasons discussed in the introduction, and also for the combination of all the four data sets, CC + Pantheon + BAO + CMB Shift.\\

From the best fit curves for $j$, one can find the deceleration parameter $q(z)$ by numerical integration. The effective equation of state parameter $w_{eff}$ is linear in $q$, so the plots for both of them will look similar. We plot $w_{eff}$ against the redshift $z$ in Fig. \ref{weffplot}. For some quoted value $q_0$ with the error bar, the upper and lower bounds of $w_{eff}$ can also be found out. The plots reveal that $w_{eff}$ has an evolution distinct from the $\Lambda$CDM model and not at all monotonically decreasing with evolution. The plots also indicate that the universe might have another stint of accelerated expansion in the recent past before entering into a decelerated phase and finally giving way to the present  accelerated expansion. \\

We started with a reconstruction of a kinematical quantity, namely the jerk parameter $j$, as this gives a flair of arriving at the evolution history without any bias towards a particular theory. As a by-product, this reconstruction leads to an evolution history of a physical quantity, the effective equation of state parameter $w_{eff}$.


\begin{thebibliography}{150}

\bibitem{perl}  S. Perlmutter {\it et al}., Astrophys. J. {\bf 517}, 565 (1999).

\bibitem{riess} A. Riess {\it et al}., Astron. J. {\bf 116}, 1009 (1998).

\bibitem{gong1} Y.G. Gong and A. Wang, Phys. Rev. D {\bf 73}, 083506 (2006).

\bibitem{gong2} Y.G. Gong and A. Wang, Phys. Rev. D {\bf 75}, 043520 (2007).

\bibitem{luongo43} O. Luongo, Mod. Phys. Lett. A \textbf{19}, 1350080 (2015).

\bibitem{rapetti44} D. Rapetti, S. W. Allen, M. A. Amin and R. D. Blandford, Mon. Not. Roy. Astron. Soc. \textbf{375}, 1510 (2007).

\bibitem{zhai45} Z.-X. Zhai, M.-J. Zhang, Z.-S. Zhang, X.-M. Liu and T.-J. Zhang, Phys. Lett. B \textbf{727}, 8 (2013).

\bibitem{ankan1} A. Mukherjee and N. Banerjee, Phys. Rev. D \textbf{93}, 043002 (2016).

\bibitem{ankan2} A. Mukherjee and N. Banerjee, Class. Quatum Grav. {\bf 34}, 03501 (2017).

\bibitem{varun} U. Alam, V. Sahni and A. A. Starobinsky, Mon. Not. R. Astron. Soc. {\bf 344}, 1057 (2003).

\bibitem{sahl1} M. Sahl\'{e}n, A. R. Liddle, and D. Parkinson, Phys. Rev. D {\bf72}, 083511 (2005).

\bibitem{sahl2} M. Sahl\'{e}n, A. R. Liddle, and D. Parkinson, Phys. Rev. D {\bf75}, 023502 (2007).

\bibitem{holsclaw1} T. Holsclaw {\it et al}., Phys. Rev. D {\bf82}, 103502 (2010).

\bibitem{holsclaw2} T. Holsclaw {\it et al}., Phys. Rev. D {\bf84}, 083501 (2011).

\bibitem{holsclaw3} T. Holsclaw {\it et al}., Phys. Rev. Lett. {\bf 105}, 241302 (2010).

\bibitem{critt} R. G. Crittenden,  G. B. Zhao,  L. Pogosian, L. Samushia and  X. Zhang, J. Cosmol. Astropart. Phys. {\bf02}, 048 (2012).

\bibitem{sanjay} R. Nair, S. Jhingan and D. Jain, J. Cosmol. Astropart. Phys. {\bf 01}, 005 (2014).

\bibitem{zzhang} Z. Zhang {\it et al}., arXiv: 1902.09794.

\bibitem{elgaroy} O. Elgaroy and T. Multamaki, Astron. Astrophys. {\bf 471}, 65 (2007).

\bibitem{carter} P. Carter, F. Beutler, W. J. Percival, J. DeRose, R. H. Wechsler and C. Zhao, Mon. Not. R. Astron. Soc. {\bf 494}, 2076 (2020).

\bibitem{rw} C. Rasmussen and C. Williams, Gaussian Processes for Machine Learning, The MIT Press (2006).

\bibitem{mackay} D. MacKay, Information Theory, Inference and Learning Algorithms, Cambridge University Press (2003), chapter 45.

\bibitem{william} C. Williams, Prediction with Gaussian processes: From linear regression to linear prediction
and beyond, in Learning in Graphical Models, ed. M. I. Jordan, 599-621. The MIT Press (1999).

\bibitem{gp} Gaussian Process webpage\\ \texttt{http://www.gaussianprocess.org/}.

\bibitem{1606.04398[24]} T. Holsclaw, U. Alam, B. Sans\'{o}, H. Lee, K. Heitmann, S. Habib, and D. Higdon, Phys. Rev. Lett. \textbf{105}, 241302 (2010).

\bibitem{1606.04398[25]} M. Seikel, C. Clarkson, and M. Smith, J. Cosmol. Astropart. Phys. \textbf{06}, 036 (2012).

\bibitem{1606.04398[26]} A. Shafieloo, A. G. Kim, and E. V. Linder, Phys. Rev. D \textbf{85}, 123530 (2012).

\bibitem{1606.04398[27]} S. Yahya, M. Seikel, C. Clarkson, R. Maartens, and M. Smith, Phys. Rev. D \textbf{89}, 023503 (2014).

\bibitem{1606.04398[28]} S. Santos-da Costa, V. C. Busti, and R. F. Holanda, J. Cosmol. Astropart. Phys. \textbf{10}, 061 (2015).

\bibitem{1606.04398[29]} T. Yang, Z.-K. Guo, and R.-G. Cai, Phys. Rev. D \textbf{91}, 123533 (2015).

\bibitem{1606.04398[30]} R.-G. Cai, Z.-K. Guo, and T. Yang, Phys. Rev. D \textbf{93}, 043517 (2016).

\bibitem{wang-meng} D. Wang, X.-H. Meng, Phys. Rev. D \textbf{95}, 023508 (2017).

\bibitem{wang-meng2} D. Wang, W. Zhang, and X.-H. Meng, Eur.Phys. J. C \textbf{79}, 211 (2019).

\bibitem{zhou-peng} L. Zhou, X. Fu, Z. Peng, J. Chen, Phys. Rev. D \textbf{100}, 123539 (2019).

\bibitem{cai-saridakis} Y.-F. Cai, M. Khurshudyan, E. N. Saridakis, Astrophys. J. \textbf{888}, 62 (2020).

\bibitem{seikel2013} M. Seikel and C. Clarkson, arXiv:1311.6678.

\bibitem{Zhang[61]} R. Jimenezand, A. Loeb, Astrophys. J. \textbf{573}, 37 (2008).

\bibitem{Zhang[62]} J. Simon, L. Verde, R. Jimenez, Phys. Rev. D \textbf{71}, 123001 (2005).

\bibitem{Zhang[63]} D. Stern, R. Jimenez, L. Verde, M. Kamionkowski, S.A. Stanford, J. Cosmol. Astropart. Phys. \textbf{2}, 008 (2010).

\bibitem{Zhang[64]} E. Gaztanaga, A. Cabre, L. Hui, Mon. Not. R. Astron. Soc. \textbf{399}, 1663 (2009).

\bibitem{Zhang[65]} M. Moresco, A. Cimatti, R. Jimenez, L. Pozzetti \textit{et al.}, J. Cosmol. Astropart. Phys. \textbf{08}, 006 (2012).

\bibitem{Zhang[66]} T. Delubac, J. Rich, S. Bailey \textit{et al.}, Astron. Astrophys. \textbf{552}, A96 (2013).

\bibitem{geng} J.-J. Geng, R.-Y. Guo, A. Wang, J.-F. Zhang, and X. Zhang, (2018), arXiv:1806.10735.

\bibitem{pan1} D. M. Scolnic, \textit{et al}., Astrophys. J. 859, 101 (2018).

\bibitem{pan2} The numerical data of the full Pantheon SNIa sample are available at- \\
\texttt{http://dx.doi.org/10.17909/T95Q4X}.\\
\texttt{https://archive.stsci.edu/prepds/ps1cosmo/index.html}.

\bibitem{planck_cmb} P. A. R. Ade \textit{et al.}, arXiv: 1502.01590. [Planck Collaboration]

\bibitem{planck} N. Aghanim \textit{et al.}, arXiv : 1807.06209. [Planck Collaboration]

\bibitem{riess1} A. G. Riess \textit{et al.}, Astrophys. J. \textbf{876}, 85 (2019).

\bibitem{q0choice} S. Capozziello, O. Farooq, O. Luongo and B. Ratra, Phys. Rev. D \textbf{90}, 044016 (2014).

\bibitem{capoz} S. Capozziello, R. D'Agostino and O. Luongo, Mon. Not. R. Astron. Soc. {\bf 494}, 2576 (2020).

\bibitem{cald} R. R. Caldwell and M. Kamionkowski,  J. Cosmol. Astropart. Phys. {\bf 0409}, 009 (2004).

\bibitem{stacho} M. P. Dabrowski and T. Stachowiak, Annals Phys. {\bf 321}, 771 (2006).



\end{thebibliography}
\end{document}